\documentclass[options]{JHEP3}

\title{TeV scale black holes thermodynamics with extra
dimensions and quantum gravity effects}

\author{Kourosh Nozari\footnote{knozari@umz.ac.ir} \quad and \quad Pegah Shahini \footnote{p.shahini@stu.umz.ac.ir} \\

Gravitation and Cosmology Group,\\ Department of Physics, Faculty of
Basic Sciences,
University of Mazandaran,\\
P. O. Box 47416-95447, Babolsar, IRAN}

\abstract{TeV scale black hole thermodynamics in the presence of
quantum gravity effects encoded in the existence of a minimal length
and a maximal momentum is studied in a model universe with large
extra dimensions.}

\begin{document}

\section{Introduction}
In the last two decades, investigations revealed that the
fundamental Planck scale in model universes with compactified Large
Extra Dimensions (LEDs) might be as small as a TeV [1]. This is a
promising feature with hope to observe TeV-scale black holes in
experiments such as the LHC. Possible production and detection of
TeV-scale black holes in colliders such as the LHC provide a
suitable basis to test our comprehending of black hole physics at
Planck scale and quantum gravity proposal. TeV scale black holes
thermodynamics in the presence of quantum gravity effects encoded in
the generalized uncertainty principle admitting a minimal measurable
length and also in loop quantum gravity-based modified dispersion
relations has been studied extensively in recent years (see for
instance [2-10]). It has been shown that the leading order
correction to the standard Bekenstein-Hawking entropy-area relation
is generally a logarithmic correction term. For extra-dimensional
gravity at TeV-scale, this leading order correction term leads to a
significant change in the possibility of formation and detection of
TeV black hole in the laboratories such as the LHC. Recently and in
the context of doubly special relativity it has been argued that a
particle's momentum cannot be arbitrary imprecise leading to the
nontrivial assumption of existence of maximal particle's momentum of
the order of the Planck momentum [11]. Existence of a maximal
particle's momentum has significant effects on the formulation of
the TeV black hole thermodynamics and changes significantly the
results of studies based on just the existence of a minimal
measurable length. Our goal here is to see what is the mutual
effects of the existence of a maximal momentum and a minimal length
on the thermodynamics of a TeV-scale black hole and how these
natural cutoffs affect the possible production of black holes in
experiments such as the LHC. Especially we focus on the role played
by the maximal momentum in this setup. With this motivation, we
consider a generalized uncertainty principle (GUP) that admits the
existence of both a minimal observable length and a maximal momentum
in a model universe with large extra dimensions. Then we study TeV
scale black hole thermodynamics in this setup. Especially the final
stage of evaporation of these black holes through Hawking radiation
is treated with details and possible relation between spacetime
dimensionality and final stage thermodynamical quantities is
explored. Finally the role played by maximal momentum on the
detectability of TeV black hole at the LHC is discussed.

\section{Minimal Length and TeV-Scale Black Hole Thermodynamics}

All approaches to quantum gravity proposal support the idea that
there is a fundamental length scale of the order of the Planck
length, which cannot be probed in a finite time. For example, in
string theory, we cannot probe distances smaller than the string
length. This minimal observable length can be realized by the
generalized uncertainty principle (GUP) as a generalization of the
standard Heisenberg uncertainty relation in the presence of
gravitational effects. Hence, the standard uncertainty principle is
replaced by the GUP that reflects quantum nature of gravity at very
short distances phenomenologically and can be formulated by a
generalized uncertainty relation as follows
\begin{equation}
\Delta x_i\geq\frac{1}{2\Delta p_i}+\alpha^{2} L_{p}^{2}\frac{\Delta
p_i}{2},
\end{equation}
where the fundamental Planck length is defined as $L_p=
G_d^{\frac{1}{d-2}}$, and $G_d$ is gravitational constant in
$d$-dimensional spacetime which in the ADD scenario takes the form
$G_d = G_4R^{d-4}$, that $R$ is the size of extra dimension(s). In
this paper natural units are used so that $\hbar=c=k_B=1$. In the
standard limit, $\Delta x\gg L_p$, one recovers the standard
uncertainty relation, $\Delta x\Delta p\geq\frac{1}{2}$. It should
be noted that the correction term in GUP formula (2.1) becomes
noticeable when momentum and distances scales are nearby the Planck
scale. In this section we use relation (2.1) to study thermodynamics
of a TeV scale Schwarzschild black hole. Our main goal here is to
see the relation between specetime dimensionality and possible
explanation of the logarithmic prefactor in black hole entropy
relation. In this manner, we deduce numerical results for
logarithmic prefactor in terms of spacetime dimensionality.

Based on the relation (2.1), a simple calculation gives
\begin{equation}
\Delta p_i\simeq\frac{\Delta x_i}{\alpha^2
L_p^2}\left(1-\sqrt{1-\frac{\alpha^2 L_p^2}{\Delta x_i^{2}}}\right).
\end{equation}
By a heuristic approach relying on Heisenberg uncertainty relation,
one deduces the following equation for Hawking temperature of black
holes [12]
\begin{equation}
T=\frac{d-3}{2\pi}\Delta p_i,
\end{equation}
which we have set the constant of proportionality equal to
$\frac{(d-3)}{2\pi}$ as a calibration factor in a $d$-dimensional
model (see [13] for instance). So, the modified black hole
temperature based on the GUP (2.1) becomes
\begin{equation}
T=\bigg(\frac{d-3}{2\pi}\bigg)\bigg(\frac{\Delta x_i}{\alpha^2
L_p^2}\bigg)\left[1-\sqrt{1-\frac{\alpha^2 L_p^2}{\Delta
x_i^{2}}}\right].
\end{equation}
In the vicinity of the black hole surface there is an inherent
uncertainty in the position of any particle of about the
Schwarzschild radius, $r_s$. So, we can set
\begin{equation}
\Delta x_i\approx r_s=\omega_dL_pm^{\frac{1}{d-3}}.
\end{equation}
In this relation, $m$ and $\omega_d$ (the dimensionless area
coefficient) are given by
\begin{equation}
\omega_d=\left(\frac{8\pi^{\frac{3-d}{2}}\Gamma\left(\frac{d-1}{2}\right)}
{d-2}\right)^{\frac{1}{d-3}}\quad,\quad m=\frac{M}{M_p}
\end{equation}
where the fundamental Planck mass is defined as
$M_p=G_d^{-\frac{1}{d-2}}=\frac{1}{L_p}$. Substituting equation
(2.5) into equation (2.4), we find
\begin{equation}
T=\bigg(\frac{d-3}{2\pi}\bigg)\bigg(\frac{\omega_d
m^{\frac{1}{d-3}}}{\alpha^2 L_p}\bigg)\left[1-\sqrt
{1-\frac{\alpha^2}{\omega_d^{2}m^{\frac{2}{d-3}}}}\right].
\end{equation}
To calculate an analytic form of black hole entropy, let us acquire
a Taylor expansion of equation (2.7) around $\alpha=0$,
\begin{equation}
T=\bigg(\frac{d-3}{2\pi}\bigg)\bigg(\frac{\omega_d
m^{\frac{1}{d-3}}}{\alpha^2
L_p}\bigg)\left[1-\left(1-\frac{\alpha^2}{2\omega_d^{2}m^{\frac{2}{d-3}}}-
\frac{\alpha^4}{8\omega_d^{4}m^{\frac{4}{d-3}}}+O(\alpha^6)\right)\right],
\end{equation}
which can be rewritten as follows
\begin{equation}
T\approx\bigg(\frac{d-3}{16\pi
L_p}\bigg)\left[\frac{4}{\omega_dm^{\frac{1}{d-3}}}+\frac{\alpha^2}{\omega_d^3m^{\frac{3}{d-3}}}\right]\,,
\end{equation}
where we have saved only terms up to the second order of $\alpha$.
Now, the related entropy can be derived by the first law of black
hole thermodynamics, that is, $dS=\frac{dM}{T}$ as,
\begin{equation}
S=\int_{M_{min}}^MdM\bigg(\frac{16\pi
L_p}{d-3}\bigg)\left[\frac{4}{\omega_dm^{\frac{1}{d-3}}}+\frac{\alpha^2}{\omega_d^3m^{\frac{3}{d-3}}}\right]^{-1},
\end{equation}
or
\begin{equation}
S=\int_{M_{min}}^Mdm\frac{16\pi
}{d-3}\left[\frac{4}{\omega_dm^{\frac{1}{d-3}}}+\frac{\alpha^2}{\omega_d^3m^{\frac{3}{d-3}}}\right]^{-1},
\end{equation}
where we have adopted the physical selection that the entropy
vanishes at $M_{min}$ where the black hole mass is minimized. With
this selection, there is no black hole radiation for masses less
than $M_{min}$. Note that black hole temperature is indeterminate
for $M<M_{min}$, where
\begin{equation}
M_{min}=\left(\frac{\alpha}{\omega_d}\right)^{d-3}M_p.
\end{equation}
Thus, Hawking evaporation process has to be stopped when the black
hole mass reaches a Planck size remnant mass. In this situation,
temperature of black hole has a maximum extremal value. We note that
detailed thermodynamics of Planck size remnant is not well-known
yet. In fact, the real thermodynamics of such an extreme system can
be even out of the realm of standard thermodynamics. It seems that
application of non-standard thermodynamics may be more reasonable at
this scale (see for instance [14] and references therein).

Black hole entropy with some arbitrary numbers of spacetime
dimensions with $\alpha=1$ can be calculated as follows

\begin{equation}
d=4\rightarrow\quad
S=16\pi\left[\frac{1}{4}M^2-\frac{1}{64}\ln(16M^2+1)-\frac{1}{16}+\frac{1}{64}\ln(5)\right],
\end{equation}
\begin{equation}
d=5\rightarrow\quad
S=8\pi\left[\frac{\sqrt{6}M^\frac{3}{2}}{9\sqrt{\pi}}-\frac{1\sqrt{6\pi
M}}{32}+\frac{3\pi\arctan(\frac{4\sqrt{6M}}{3\sqrt{\pi}})}{128}+\frac{\sqrt{144}\pi}{192}+\frac{3\pi}{64}-\frac{3\pi\arctan(2)}{128}\right]
\end{equation}

$$d=6\rightarrow\quad
S=\frac{16\pi}{3}\bigg[\frac{3}{32}(\frac{12M^4}{\pi})^\frac{1}{3}-\frac{1}{32}(18\pi
M^2)^\frac{1}{3}+\frac{\pi}{64}\ln\bigg((144M^2)^\frac{1}{3}+(\pi^2)^\frac{1}{3}\bigg)$$
\begin{equation}
-\frac{\pi}{16}-\frac{\pi}{64}\ln\bigg((125\pi^2)^\frac{1}{3}\bigg)\bigg]
\end{equation}

$$d=7\rightarrow
S=4\pi\bigg[\frac{2M}{5}(\frac{M}{5\pi^2})^\frac{1}{4}-\frac{1}{24}(5M^3\pi^2)^\frac{1}{4}+\frac{\pi}{128}(125M\pi^2)^\frac{1}{4}-
\frac{5\pi^2}{512}\arctan(\frac{16M}{5\pi^2})^\frac{1}{4}$$
\begin{equation}
-\frac{\pi^2}{16}+\frac{5\pi^2}{192}-\frac{5\pi^2}{256}+\frac{5\pi^2}{512}\arctan(2)\bigg]
\end{equation}
and so on. From a loop quantum gravity viewpoint, the leading order
correction term to the Bekenstein-Hawking entropy of black hole has
logarithmic nature [15]. Here we see that the logarithmic term is
present just for even number of spacetime dimensionality (see also
[16]).

The heat capacity of the black hole can also be achieved utilizing
the following standard formula
\begin{equation}
C=\frac{dM}{dT}.
\end{equation}
With $\alpha=1$, we find
\begin{equation}
C=(-16m\pi)\left[\frac{4}{\omega_dm^\frac{1}{d-3}}+\frac{3}{\omega_d^3m^\frac{3}{d-3}}\right]^{-1}
\end{equation}
Therefore, we were able to calculate some important thermodynamical
quantities attributed to TeV black holes in a model universe with
large extra dimensions where quantum gravity effects are taken into
account through a generalized uncertainty principle that admits just
a minimal measurable length. The physics of TeV scale black hole
evaporation in the presence of quantum gravity effects, especially
existence of a minimal measurable length, powerfully support the
idea that the final stage of these black holes evaporation is a
stable, Planck-size remnant. These remnants may be a reliable
candidate for dark matter.

Figures 1, 2 and 3 demonstrate some features of TeV black hole
thermodynamics versus mass. In the minimal length GUP framework,
evaporation of black hole stops once it reaches the stable
Planck-size remnant and its entropy and the heat capacity in this
stage are vanishing, but its temperature reaches a maximal value.
So, evaporation of TeV scale black hole ends once black hole mass
reaches a Planck-size remnant mass. The existence of this remnant
has been considered as a possible solution to the information loss
problem in black hole evaporation. It has been suggested also that
these remnants may be a possible candidates for dark matter
[7,16,17].

\begin{figure}[htp]
\begin{center}
\includegraphics{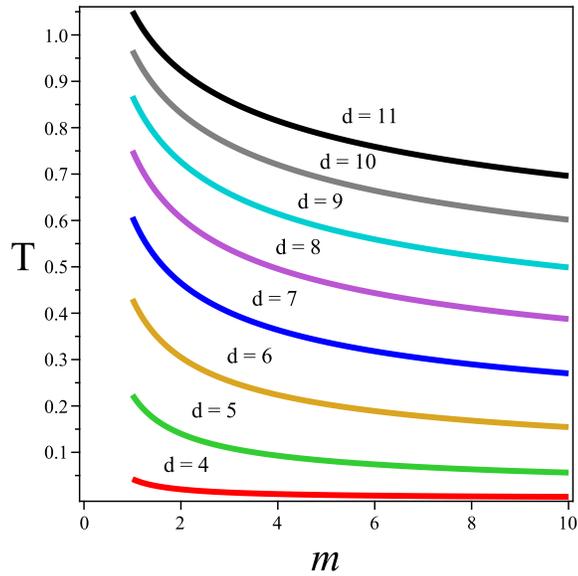} \vspace{1cm}
\end{center}
\vspace{8cm} \caption{\scriptsize{Temperature of black hole versus
its mass in different spacetime dimensions. Mass is in the units of
the Planck mass. }}
\end{figure}

\begin{figure}[htp]
\begin{center}
\includegraphics{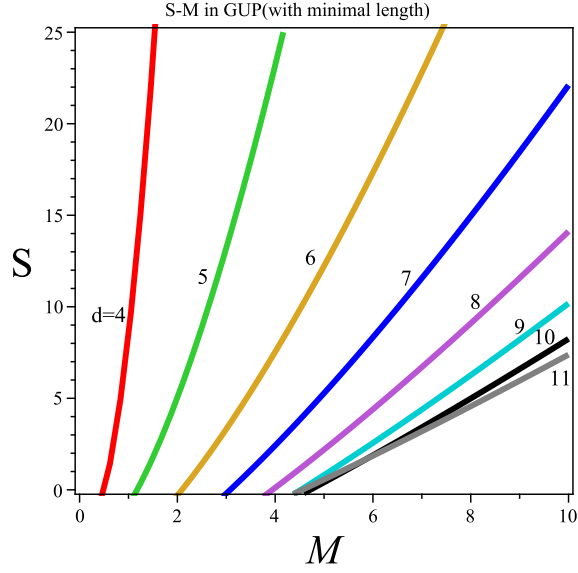} \vspace{1cm}
\end{center}
\vspace{7cm} \caption{\scriptsize{ Entropy of black hole versus its
mass in different spacetime dimensions. }}
\end{figure}

\begin{figure}[htp]
\begin{center}
\includegraphics{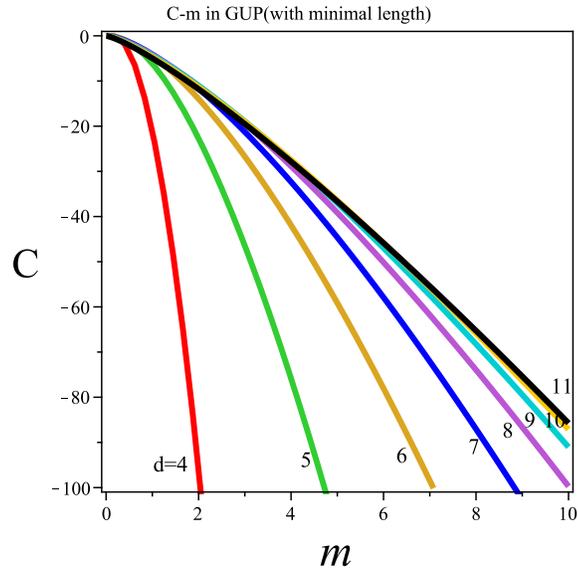} \vspace{1cm}
\end{center}
\vspace{7cm} \caption{\scriptsize{ Heat Capacity of black hole
versus its mass in different spacetime dimensions.}}
\end{figure}

\newpage

In which follows we firstly present a discussion on the issue of
maximal momentum and then we reconsider the issue of TeV black hole
thermodynamics in a model universe with large extra dimensions and
in the presence of the quantum gravity effect through a GUP that
admits both a minimal observable length and a maximal particle's
momentum. We focus mainly on the role played by the maximal momentum
and we comment on the production of TeV scale black hole at the LHC
in the presence of these quantum gravity effects. We also compare
our results with those results that are obtained by excluding the
possibility of having a natural cutoff on particle's momentum.

\section{A GUP with minimal length and Maximal momentum}

In the context of the Doubly Special Relativity (DSR), one can show
that a test particles' momentum cannot be arbitrarily imprecise. In
fact there is an upper bound for momentum fluctuations [11]. As a
nontrivial assumption, this may lead to a maximal measurable
momentum for a test particle [18]. In this framework, the GUP that
predicts both a minimal observable length and a maximal momentum can
be written as follows [18]
\begin{equation}
\Delta x\Delta p\geq\frac{1}{2}\left[1+(\frac{\alpha}{\sqrt{\langle
p^2\rangle}}+4\alpha^2)\Delta p^2+4\alpha^2\langle
p\rangle^2-2\alpha\sqrt{\langle p^2\rangle}\right].
\end{equation}
Since $(\Delta p)^2=\langle p^2\rangle-\langle p\rangle^2$,  by
setting $\langle p\rangle=0$ for simplicity, we find
\begin{equation}
\Delta x\Delta p\geq\frac{1}{2}\left[1-\alpha(\Delta
p)+4\alpha^2(\Delta p)^2\right].
\end{equation}
This GUP that contains both minimal length and maximal momentum is
the primary input of our forthcoming arguments. Before treating our
main problem, first we show how maximal momentum arises in this
setup. The absolute minimal measurable length in our setup is given
by  $\Delta x_{min}(\langle p\rangle=0)\equiv\Delta
x_0=\frac{3\alpha}{2} $.  Due to duality of position and momentum
operators, it is reasonable to assume $\Delta x_{min}\propto\Delta
p_{max}$.  Now, saturating the inequality in relation (3.2), we find
\begin{equation}
2(\Delta x\Delta p)=\bigg(1-\alpha(\Delta p)+4\alpha^2(\Delta
p)^2\bigg).
\end{equation}
This results in
\begin{equation}
(\Delta p)^2-\frac{(2\Delta x+\alpha)}{4\alpha^2}\Delta
p+\frac{1}{4\alpha^2}=0.
\end{equation}
So we find
\begin{equation}
(\Delta p_{max})^2-\frac{(2\Delta x_{min}+\alpha)}{4\alpha^2}\Delta
p_{max}+\frac{1}{4\alpha^2}=0.
\end{equation}
Now using the value of $\Delta x_{min}$, we find
\begin{equation}
(\Delta p_{max})^2-\frac{1}{\alpha}\Delta
p_{max}+\frac{1}{4\alpha^2}=0.
\end{equation}
The solution of this equation is
\begin{equation}
\Delta p_{max} =\frac{1}{2\alpha}.
\end{equation}
so, there is an upper bound on particle's momentum uncertainty. As a
nontrivial assumption, we assume that this maximal uncertainty in
particle's momentum is indeed the maximal measurable momentum. This
is of the order of Planck momentum.

\section{TeV-Scale Black Hole Thermodynamics with Natural Cutoffs}

Now we focus on the TeV-scale black hole thermodynamics in the
presence of both a minimal measurable length and a maximal momentum
as phenomenological, natural cutoffs. By assuming a model universe
with $d$ dimensions and based on the relation (3.2), by a simple
calculation we find
\begin{equation}
\Delta p_i\simeq\frac{(\Delta
x_i+\gamma)}{4\alpha^2}\left[1-\sqrt{1-\frac{4\alpha^2}{(\Delta
x_i+\gamma)^{2}}}\right],
\end{equation}
where $\gamma\equiv\frac{\alpha}{2}$. Since
$T=\frac{d-3}{2\pi}\Delta p_i$, we find
\begin{equation}
T=\bigg(\frac{d-3}{2\pi}\bigg)\bigg(\frac{\Delta
x_i+\gamma}{4\alpha^2}\bigg)\left[1-\sqrt{1-\frac{4\alpha^2}{(\Delta
x_i+\gamma)^{2}}}\right]
\end{equation}
Once again, we assume that in the vicinity of the black hole horizon
surface, the inherent uncertainty in the position of any particle is
of the order of Schwarzschild  black hole radius, $ r_s$. So we have
$\Delta x_i\approx r_s=\omega_dL_pm^{\frac{1}{d-3}}$.  Using this
result we find
\begin{equation}
T=\bigg(\frac{d-3}{2\pi}\bigg)\bigg(\frac{(\omega_d
m^{\frac{1}{d-3}}+B)}{4\alpha_0^2 L_p}\bigg)\left[1-\sqrt
{1-\frac{4\alpha_0^2}{(\omega_dm^{\frac{1}{d-3}}+B)^2}}\right]
\end{equation}
where $\alpha=\alpha_{0}L_{p}$ and $B\equiv\frac{\alpha_0}{2}$. To
obtain an analytic form for black hole entropy, we expand (4.3) in a
Taylor series about $\alpha=0$. This gives the following result up
to ${\cal{O}}(\alpha^{2})$
\begin{equation}
T=\bigg(\frac{d-3}{2\pi}\bigg)\bigg(\frac{(\omega_d
m^{\frac{1}{d-3}}+B)}{4\alpha_0^2
L_p}\bigg)\left[1-\bigg(1-\frac{4\alpha_0^2}{2(\omega_dm^{\frac{1}{d-3}}+B)^2}-
\frac{(4\alpha_0^2)^2}{8(\omega_dm^{\frac{1}{d-3}}+B)^4}\bigg)\right]\,,
\end{equation}
which can be rewritten as
\begin{equation}
T=\bigg(\frac{d-3}{16\pi
L_p}\bigg)\left[\frac{4}{(\omega_dm^{\frac{1}{d-3}}+B)}+\frac{A}{(\omega_dm^{\frac{1}{d-3}}+B)^3}\right].
\end{equation}
where we have set $A\equiv4\alpha_0^2$. This result gives the
modified black hole temperature based on GUP (3.2). To compute
entropy of TeV black hole in this setup, we use the first law of
classical black hole thermodynamics $ds=\frac{dM}{T}$ but now with
modified temperature. We find
\begin{equation}
S=\int_{M_{min}}^Mdm\bigg(\frac{16\pi }{d-3}\bigg)
\left[\frac{4}{(\omega_dm^{\frac{1}{d-3}}+B)}+\frac{A}{(\omega_dm^{\frac{1}{d-3}}+B)^3}\right]^{-1},
\end{equation}
In the same way as the previous section, we adopt the physical
selection that the evaporation process terminates when the mass of
the radiating black hole reaches a Planck-size remnant of mass
$M_{min}$. In this phase the entropy of black hole vanishes while
its temperature reaches a maximum, finite value. In which follows we
present the entropy of TeV black hole for some spacetime dimensions
for $\alpha=1$
\begin{equation}
d=4\rightarrow\quad
S=16\pi\left[{\frac{1}{4}}M^2+{\frac{1}{8}}M-{\frac{1}{16}}\ln(16M^2+8M+5)-\frac{1}{8}+{\frac{1}{16}}\ln(13)\right]
\end{equation}

$$d=5\rightarrow\quad
S=8\pi\bigg[{\frac{\sqrt{6}M^\frac{3}{2}}{9\sqrt{\pi}}}
+{\frac{1}{8}}M-{\frac{1}{8}}\sqrt{6\pi M} +{\frac{3}{64}}\pi
ln(32M+8\sqrt{6\pi M}+15\pi)$$
$$+{\frac{3}{16}}\pi\arctan\big(\frac{(64\sqrt{M}+8\sqrt{6\pi})\sqrt{6}}{96\sqrt{\pi}}\big)
-\frac{\pi\sqrt{144}}{192}$$
\begin{equation}
+\frac{9\pi}{64}-{\frac{3\pi}{64}}ln(39\pi)
-{\frac{3\pi}{16}}\arctan(\frac{3}{2}) \bigg]
\end{equation}

$$d=6\rightarrow\quad \quad \quad
S=\frac{16\pi}{3}\Bigg[\frac{3}{32}(\frac{12M^4}{\pi})^\frac{1}{3}+\frac{M}{8}-\frac{1}{8}(18\pi
M^2)^\frac{1}{3}+\frac{1}{8}(12\pi^2 M)^\frac{1}{3}$$
$$+\frac{3\pi}{16}ln\bigg((\frac{144M^2}{\pi^2})^\frac{1}{3}+(\frac{96M}{\pi})^\frac{1}{3}+5\bigg)-
\frac{\pi}{2}\arctan(\frac{(8M)^\frac{1}{3}+(\frac{2\pi}{3})^\frac{1}{3}
}{2\sqrt{2}})$$
\begin{equation}
-\frac{3}{8}+\frac{\pi}{6}-\frac{3\pi}{6}ln(13)+\frac{\pi}{2}\arctan(\frac{(108)^\frac{1}{3}}{2\sqrt{2}})\Bigg]
\end{equation}

$$d=7\rightarrow\quad\quad\quad
S=4\pi\Bigg[\frac{2}{25}(\frac{125M^5}{\pi^2})^\frac{1}{4}+\frac{M}{8}-\frac{1}{6}(5\pi^2M^3)^\frac{1}{4}+\frac{1}{16}\sqrt{5\pi^2M}
$$ $$+\frac{3}{32}(125\pi^6M)^\frac{1}{4}-\frac{55\pi^2}{256}ln\bigg(\frac{16\sqrt{5M}}{\pi}+8(\frac{125M}{\pi^2})^\frac{1}{4}+25\bigg)-\frac{71\pi^2}{384}
$$
\begin{equation}
-\frac{5\pi^2}{64}\arctan\bigg(\frac{4(\frac{125M}{\pi^2})^\frac{1}{4}+5}{10}\bigg)+\frac{55\pi^2}{256}ln(65)+\frac{5\pi^2}{64}\arctan(\frac{1}{2})\Bigg]
\end{equation}

and so on.

The heat capacity of black hole can be computed by using the
relation $C=\frac{dM}{dT_H}$
\begin{equation}
\alpha=1\rightarrow
C=\frac{-16\pi}{\omega_d}m^\frac{d-4}{d-3}\left[\frac{4}{(\omega_dm^\frac{1}{d-3}+\frac{1}{2})^2}+\frac{12}{(\omega_dm^\frac{1}{d-3}+\frac{1}{2})^4}\right]^{-1}
\end{equation}

Hence we were able to compute some important thermodynamical
properties of black holes by utilizing a new version of GUP
admitting both a minimal length and a maximal momentum. The main
ingredient of this analysis is the existence of a stable,
Planck-size remnant.  Figures 4, 5 and 6 show the temperature,
entropy and heat capacity of the black hole versus its mass in the
presence of both minimal length and maximal momentum. Similar to the
previous section, the evaporation process of black hole ends up in a
stable, Planck-size remnant with vanishing entropy and heat capacity
but its temperature reaches a maximal value. We focus on the role
played by the natural cutoff on momentum in the next section.

\begin{figure}[htp]
\begin{center}
\includegraphics{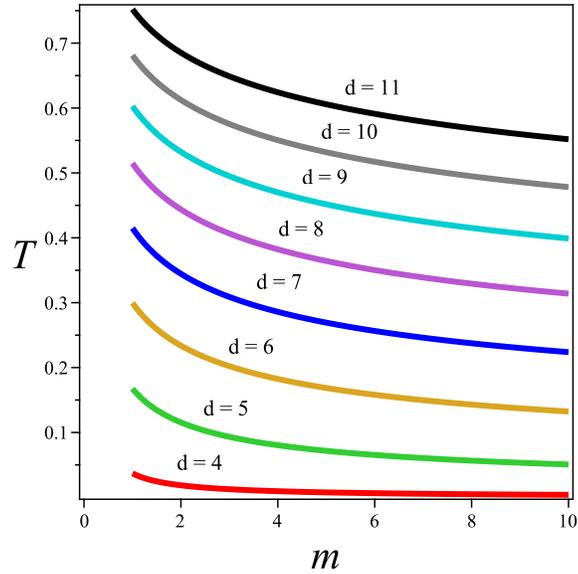} \vspace{1cm}
\end{center}
\vspace{7cm} \caption{\scriptsize{Temperature of black hole versus
its mass in the presence of both minimal length and maximal momentum
in different spacetime dimensions. Mass is in the unit of the Planck
mass. }}
\end{figure}

\begin{figure}[htp]
\begin{center}
\includegraphics{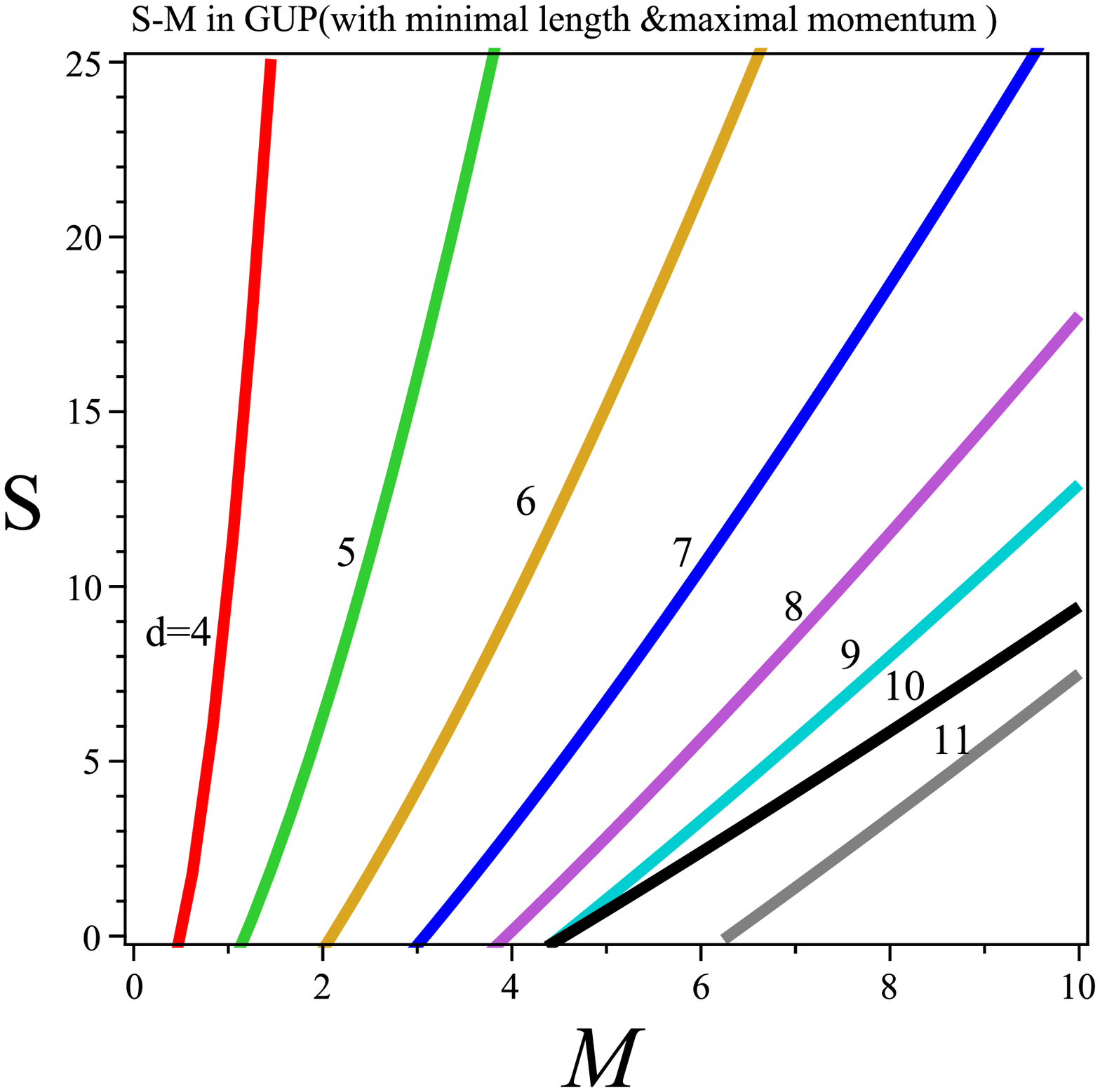} \vspace{1cm}
\end{center}
\vspace{7.5cm} \caption{\scriptsize{ Entropy of black hole versus
its mass in the presence of both minimal length and maximal momentum
in different spacetime dimensions. }}
\end{figure}

\begin{figure}[htp]
\begin{center}
\includegraphics{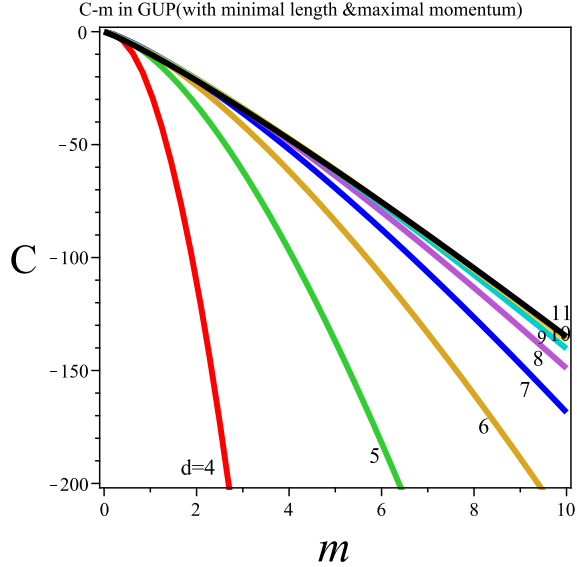} \vspace{1cm}
\end{center}
\vspace{6cm} \caption{\scriptsize{ Heat Capacity of black hole
versus its mass in the presence of both minimal length and maximal
momentum in different spacetime dimensions.}}
\end{figure}

\newpage
\section{Discussion and Results}
The final stage of a TeV scale black hole evaporation in the
presence of quantum gravity effect encoded in the existence of a
minimal measurable length is a stable remnant. In this phase the
entropy and heat capacity are zero, but the temperature reaches a
maximal value that depends explicitly on the spacetime
dimensionality. This temperature increases when the dimensionality
of spacetime increases. When we incorporate also the Doubly Special
Relativity motivated maximal momentum as a natural cutoff on a test
particle's momentum, the overall behavior of thermodynamical
quantities are the same as the case with just a minimal length, but
now the maximum value of temperature in final stage of evaporation
is less than the case with just a minimal length cutoff. This is
physically reasonable since existence of high momentum cutoff
suppresses the contribution of highly excited states that were not
forbidden in ordinary situation. The black hole remnant in a model
universe with large extra dimensions is hotter than its
4-dimensional counterpart. On the other hand, by increasing the
number of extra dimensions, the minimum mass of black hole remnant
increases. This means that possibility of creation and detection of
black holes in the LHC or any high energy physics laboratory
decreases by increasing the number of spacetime dimensions. This is
because the minimum energy for creation of black hole in extra
dimensions increases.

\begin{figure}[htp]
\begin{center}
\includegraphics{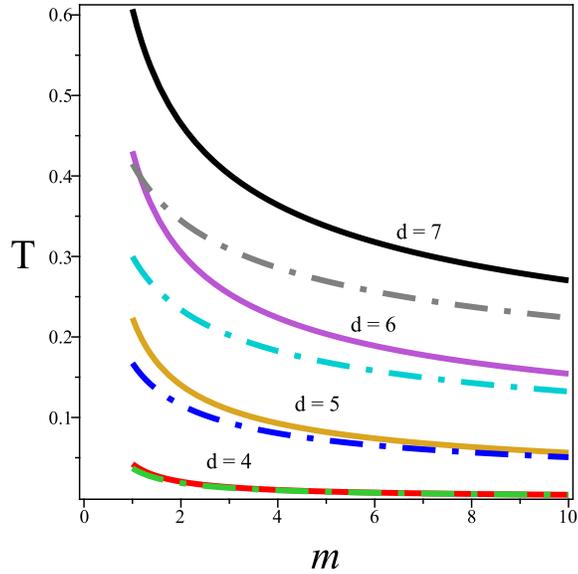} \vspace{1cm}
\end{center}
\vspace{6cm} \caption{\scriptsize{Temperature of black hole versus
its mass in different spacetime dimensions. Mass is in the unit of
the Planck mass (solid lines for minimal length GUP and dashed-lines
for minimal length and maximal momentum GUP).}}
\end{figure}

\begin{figure}[htp]
\begin{center}
\includegraphics{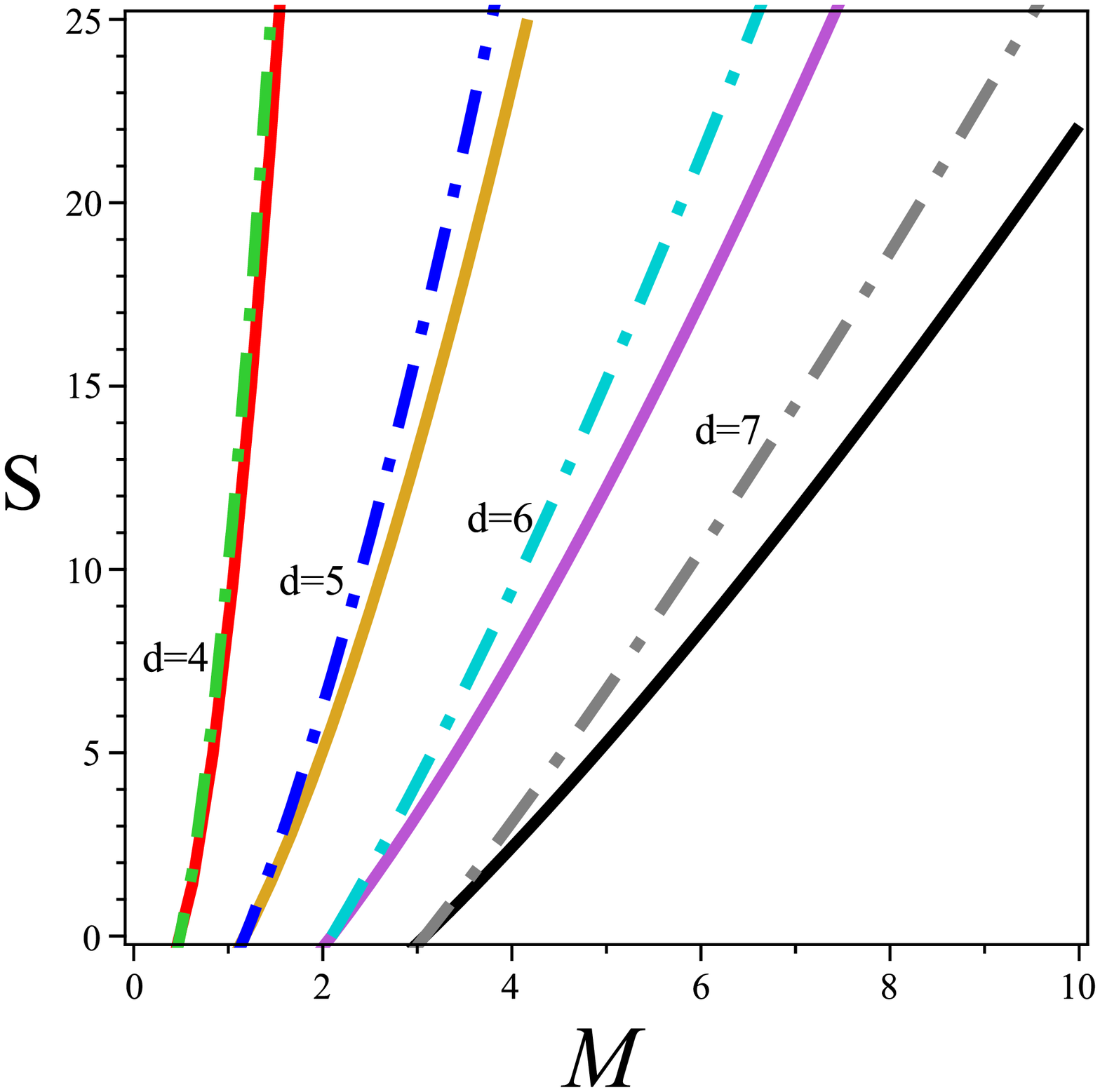} \vspace{1cm}
\end{center}
\vspace{7cm} \caption{\scriptsize{ Entropy of black hole versus its
mass in different spacetime dimensions (solid lines for minimal
length GUP and dashed-lines for minimal length and maximal momentum
GUP). }}
\end{figure}

\begin{figure}[htp]
\begin{center}
\includegraphics{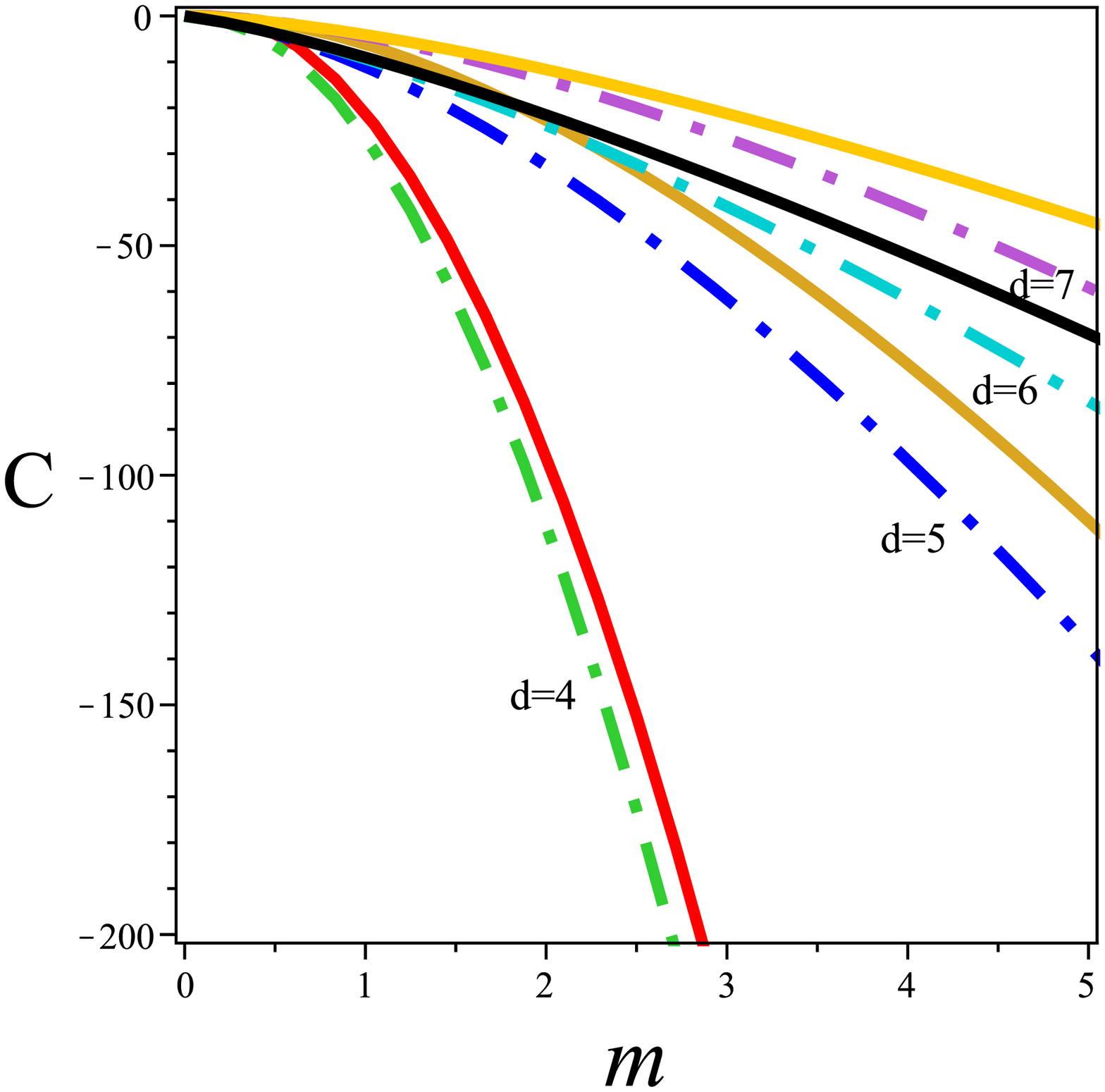} \vspace{1cm}
\end{center}
\vspace{6cm} \caption{\scriptsize{ Heat Capacity of black hole
versus its mass in different spacetime dimensions (solid lines for
minimal length GUP and dashed-lines for minimal length and maximal
momentum GUP).}}
\end{figure}

In figures 7, 8 and 9 we compare thermodynamical quantities of the
black hole computed once with just a minimal length and in the other
case with both minimal length and maximal momentum to highlight the
role played by natural cutoff on particle's momentum. Figure 7 shows
that the temperature of black hole decreases by considering the
maximal momentum. Figure 8 shows that by increasing the number of
spacetime dimensionality, the minimum mass increases. Also this
figure shows that as a result of natural cutoff on momentum, entropy
of the black hole increases relative to the case that this cutoff is
ignored. As figure 9 shows, the heat capacity of black hole
increases in magnitude when we consider the effect of momentum
natural cutoff. In summary, when we consider both minimal length and
maximal momentum, $M_{min}$ is smaller than the case that we
consider just the minimal length. This means that possibility of
creation and detection of black holes at the LHC increases in this
situation. We note that all previous studies in this direction were
neglected the possible existence of a natural cutoff on a test
particle's momentum. Here we considered this phenomenological aspect
of quantum gravity and we have shown that it leads to more probable
creation and detection of TeV scale black holes at the LHC.Tables
$1$ and $2$ summarize our results for $\alpha=1$ and
$\alpha=\frac{2}{\pi}$.

\begin{table}
\caption{{\small GUP-corrected maximum temperature and minimum mass
of black hole for $\alpha=1$ in scenarios with large extra
dimensions. Mass is in units of the Planck mass and temperature is
in units of the Planck energy (supposing $M_p=1$\,TeV).}}
\begin{center}
\begin{tabular}{|c||c|c|c|}
\hline& $M_{min}^{{\tiny \alpha=1}}(TeV)$ & $T_{max}^{{\tiny
\alpha=1}}(TeV)(min)$ & $T_{max}^{{\tiny
\alpha=1}}(TeV)(minmax)$\\
\hline$d=4$ & $0.5$ & $0.1$ & $0.08$ \\
\hline$d=5$ & $1.18$ & $0.2$ & $0.15$ \\
\hline$d=6$ & $2.09$ & $0.3$ & $0.23$ \\
\hline$d=7$ & $3.08$ & $0.4$ & $0.31$ \\
\hline$d=8$ & $3.94$ & $0.5$ & $0.38$\\
\hline$d=9$ & $4.51$ & $0.6$ & $0.46$ \\
\hline$d=10$ & $4.72$ & $0.7$ & $0.54$ \\
\hline$d=11$ & $4.56$ & $0.8$ & $0.61$\\
\hline
\end{tabular}
\end{center}
min: GUP with just a minimal length\\
minmax: GUP with both minimal length and maximal momentum
\end{table}

\begin{table}
\caption{{\small GUP-corrected maximum temperature and minimum mass
of black hole for $\alpha=\frac{2}{\pi}$ in scenarios with large
extra dimensions. Mass is in units of the Planck mass and
temperature is in units of the Planck energy (supposing
$M_p=1$\,TeV).}}
\begin{center}
\begin{tabular}{|c||c|c|c|}
\hline& $M_{min}^{{\tiny \alpha=\frac{2}{\pi}}}(TeV)$ &
$T_{max}^{{\tiny \alpha=\frac{2}{\pi}}}(TeV)(min)$ &
$T_{max}^{{\tiny
\alpha=\frac{2}{\pi}}}(TeV)(minmax)$\\
\hline$d=4$ & $0.32$ & $0.16$ & $0.12$ \\
\hline$d=5$ & $0.48$ & $0.31$ & $0.25$ \\
\hline$d=6$ & $0.54$ & $0.47$ & $0.37$ \\
\hline$d=7$ & $0.51$ & $0.62$ & $0.5$ \\
\hline$d=8$ & $0.41$ & $0.78$ & $0.62$\\
\hline$d=9$ & $0.3$ & $0.94$ & $0.74$ \\
\hline$d=10$ & $0.2$ & $1.09$ & $0.87$ \\
\hline$d=11$ & $0.12$ & $1.25$ & $0.99$\\
\hline
\end{tabular}
\end{center}
\end{table}


\newpage
\begin{thebibliography}{10}
\bibitem{1}
N. Arkani-Hamed, S. Dimopoulos and G. Dvali, Phys. Lett. B
\textbf{429}, 263 (1998)\\
I. Antoniadis, N. Arkani-Hamed, S. Dimopoulos and G. Dvali, Phys.
Lett. B \textbf{436}, 257 (1998)\\
L. Randall and R. Sundrum, Phys. Rev. Lett. \textbf{83}, 3370
(1999).

\bibitem{2}
K. Nozari and A. S. Sefiedgar, Gen. Relat. Gravit. {\bf39}, 501
(2007).

\bibitem{3}
A. J. M. Medved, Class. Quant. Grav. {\bf22}, 133 (2005)\\
G. Gour and A. J. M. Medved, Class. Quant. Grav. {\bf20}, 3307
(2003)\\
A. J. M. Medved, Class. Quant. Grav. {\bf20}, 2147 (2003).

\bibitem{4}
R. J. Adler {\it et al}, Gen. Rel. Grav. {\bf 33}, 2101 (2001).

\bibitem{5}
K. Nozari and S. H. Mehdipour, Mod. Phys. Lett. A {\bf20}, 2937
(2005).

\bibitem{6}
G. Amelino-Camelia, M. Arzano, Y. Ling and G. Mandanici, Class.
Quant. Grav. \textbf{23}, 2585 (2006).

\bibitem{7}
K. Nozari and S. H. Mehdipour, Europhys. Lett. \textbf{84}, 20008
(2008)\\
K. Nozari and S. H. Mehdipour, Class. Quant. Grav. \textbf{25},
175015 (2008).

\bibitem{8}
W. Kim, E. J. Son and M. Yoon, JHEP \textbf{0801}, 035 (2008).

\bibitem{9}
L. Xiang and X. Q. Wen, JHEP \textbf{0910}, 046 (2009).

\bibitem{10}
R. Banerjee and S. Ghosh, Phys. Lett. B, \textbf{688}, 224 (2010).

\bibitem{11}
J. Magueijo and L. Smolin, Phys. Rev. Lett. \textbf{88}, 190403
(2002)\\
J. Magueijo and L. Smolin, Phys. Rev. Lett. D \textbf{ 67}, 044017
(2003)\\
J. Magueijo and L. Smolin, Phys. Rev. D \textbf{71}, 026010 (2005)\\
J. L. Cortes and J. Gamboa, Phys. Rev. D \textbf{71}, 065015 (2005).

\bibitem{12}
H. Ohanian and R. Ruffini, \emph{Gravitation and
Spacetime}, 2nd edition (W. W. Norton) 1994, p. 481.

\bibitem{13}
M. Cavagli\`a and S. Das, Class. Quant. Grav. {\bf21}, 4511 (2004)\\
M. Cavagli\`a, {\it et al}, Class. Quant. Grav. {\bf 20}, L205
(2003).

\bibitem{14}
K. Nozari and S. H. Mehdipour, Chaos Solitons and Fractals,
\textbf{39}, 956 (2009).

\bibitem{15}
A. J. M. Medved, Elias C. Vagenas, Phys. Rev. D \textbf{70}, 124021
(2004).

\bibitem{16}
K. Nozari and A. S. Sefiedgar, Gen. Relat. Gravit. \textbf{39}, 501
(2007)\\
K. Nozari, Astrophys. Space Science, DOI 10.1007/s10509-012-1078-6

\bibitem{17}
K. Nozari, Astropart. Phys. \textbf{27}, 169 (2007).

\bibitem{18}
A. F. Ali, S. Das and E. C. Vagenas, Phys. Rev. D {\textbf{84}},
044013 (2011)\\
P. Pedram, K. Nozari and S. H. Taheri,  JHEP \textbf{1103}, 093
(2011)\\
K. Nozari and A. Etemadi, Phys. Rev. D \textbf{85}, 104029 (2012)
[arXiv:1205.0158].




\end{thebibliography}
\end{document}